\documentstyle[12pt]{article}
\begin{document}

\begin{flushright}{CMU-HEP95-04\\
DOE-ER/40682-93}
\end{flushright}

\vspace{.25in}

\centerline{\bf CP Violation in $D^0-overline{D^0}$ Mixing}

\vspace{.25in}

\begin{center}
{Lincoln Wolfenstein\\
Carnegie Mellon University\\
Department of Physics\\
Pittsburgh, PA  15213}
\end{center}

\vspace{.25in}

\begin{abstract}
The existence of $D^0-\overline{D^0}$ mixing at a detectable level requires new
physics, which effectively yields a $\Delta c = 2$ superweak interaction.  In
general this interaction may involve significant CP violation.  For small
values of the mixing it may be much easier to detect the CP-violating part of
the mixing than the CP-conserving part.
\end{abstract}

$D^0-\overline{D^0}$ mixing is expected to be very small in the standard
model.{\cite{1}}  While quantitative estimates are difficult because of
long-distance effects,{\cite 2\cite3} it is clear that
$x \left(\equiv \Delta m_D/\Gamma_D\right)$ is well below $10^{-2}$ whereas the
present limit is .06.  Future experiments that could probe for a value of $x$
around $10^{-2}$ would therefore be a way of discovering new physics.

Many extensions of the standard model can lead to $D^0-\overline{D^0}$ mixing
with
$x \sim 10^{-2}$.  Examples include models with an extra $Q= 2/3$ isosinglet
quark{\cite{4}} and the general two Higgs doublet model (2HDM) with
flavor-changing
neutral exchange (FCNE).{\cite{5}}  In these models it is quite natural that
there is a
significant CP violation associated with the mixing.  Here we point out that
the
study of the time dependence{\cite{6}} of the decay $D^0\rightarrow K^+ \pi^-$
is
particularly sensitive to the CP-violating part of the mixing.  Indeed it may
be
much easier to detect  $D^0-\overline{D^0}$ mixing if it has a sizeable CP
violation{\cite{7}} than if it is approximately CP-conserving.

In addition to the dependence on $\Delta m$, the time evolution of a $D^0$ beam
depends on $\Delta \Gamma$.  We assume here that the new physics does not
affect the
decays significantly and therefore does not affect $\Delta \Gamma$; rather the
new
physics produces an effective superweak $\Delta c = 2$ interaction.  However in
the
standard model $\Delta \Gamma/\Gamma$ is small{\cite{2}\cite{3}} for the same
reasons as $\Delta m /\Gamma$; indeed $\Delta \Gamma$ is just the absorptive
part of the long-distance diagrams that determine $\Delta m$.  Thus we neglect
$\Delta \Gamma$.

Including the effects of CP violation the CP eigenstates $D_1 \left(\equiv
D^0+\overline{D^0}\right)$ and $D_2\left(\equiv  D^0-\overline{D^0}\right)$ are
related to the mass eigenstates $D_H$ and $D_L$ by

$$D_1 = \cos \phi D_H + i \sin \phi D_L$$
$$D_2 = i \sin \phi D_H + \cos \phi D_L\eqno{(1)}$$

The factor $i$ is a consequence of CPT invariance when $\Delta \Gamma << \Delta
m$
just as in the case{\cite{8}} of the $B^0 - \overline{B^0}$ system.  Then as a
function of time the state starting as a $D^0$ evolves with time as

$$e^{\tau} D^0 \left(\tau \right) = D^0 \cos x\tau + e^{2i \phi} \overline{D^0}
\left(-i \sin x \tau \right)\eqno{(2)}$$

\noindent where $\tau = t \Gamma/2$.

The $D^0$ can decay to $K^+ \pi^-$ via the doubly Cabibbo suppressed
amplitude (DCSD) ${\cal E}Ae^{i \delta_D}$ where ${\cal E} \simeq
\tan^2\theta_c
\simeq .05.$  In contrast the allowed decay $\overline{D^0}$ has the amplitude
$Ae^{i
\delta_A}$.  Here $\delta_A$ and $\delta_D$ are the ``strong'' phase shifts.
It
might seem obvious that $\delta_A = \delta_D$ since we are dealing with the
same
final state $K^+\pi^-$.  This would be true if there were only elastic
scattering.
In fact the phases $\delta_D$ and $\delta_A$ must be derived from the
absorptive part
of the amplitudes for $D^0 \rightarrow K^+ \pi^-$ and $\overline{D^0}
\rightarrow
K^+ \pi^-$, respectively.  These are not the same since the effective weak
operators
leading to these decays are different.  However, one can show that $\delta_A =
\delta_D$ in the SU(3) limit.  By CP invariance, which holds to a very good
approximation for the weak decay assumed to be governed by the standard model,
the
final state phase for $D^0 \rightarrow K^- \pi^+$ is given by $\delta_A$.  By
the
interchange of the quarks $s$ and $d$ in the effective operators and in the
final
state the allowed decay amplitude $D^0 \rightarrow K^- \pi^+$ becomes the DCSD
amplitude $D^0 \rightarrow K^+ \pi^-$, so that in this SU(3) approximation
$\delta_A
= \delta_D$.

With this assumption and letting $x \tau << 1$ the decay rate has the time
dependence

$$R\left(K^+\pi^-\right) = e^{-2 \tau}A^2{\cal E}^2\left\{1+2\left[\sin\left(2
\phi\right)\right]\left(\frac{x\tau}{{\cal E}}\right) + \left(\frac{x^2
\tau^2}{{\cal
E}^2}\right)\right\} \eqno{(3)}$$

For an initial $\overline{D^0}$ state going to $K^- \pi^+$ the sign of the $x
\tau$
term is reversed.  We now see that the linear term in time which is most
sensitive
to small values of $x$ occurs only in the case of CP-violating
$D^0-\overline{D^0}$
mixing.  For example, for $x \tau = .01$, the CP-violating term is of order
40\%
for $\sin 2\phi = 1$, whereas the quardratic term is only 4\%.  Of course, the
CP
violation can be directly detected by comparing the $D^0$ and $\overline{D^0}$
decays.  If the quadratic term is not detected one can measure only the product
$x
\sin 2 \phi$, the CP-violating part of the mixing matrix and not $x$ and $\sin
2
\phi$ separately.

With these assumptions it is not necessary to measure the time distribution.
The
difference between the integrated rate for $D^0 \rightarrow K^+ \pi^-$ and that
for
$\overline{D^0} \rightarrow K^- \pi^+$ directly measures $x \sin 2 \phi$ and
reveals both mixing and CP violation.  Without our assumptions, of course, this
difference could be due to CP-violation in some new physics contribution to the
decay
amplitude.  However, whereas there are many models that suggest
$D^0-\overline{D^0}$ mixing, it is hard to find one that contributes
significantly to
the decay amplitude.

The general reason that CP violation is important is that  $D^0-\overline{D^0}$
mixing depends on $\left(\Delta m t\right)$, or $x \tau$, and for small values
of
$x$ one is most sensitive to the linear term in $x \tau$.  This term is odd
under
the change of $t$ to $-t$.  In the \underline{absence} of ``final state
interactions,'' that is, of absorptive parts of diagrams, this corresponds to
time
reversal.  Thus once we set $\Delta \Gamma$ and $\delta_A - \delta_B$ to zero
the term proportional to $x \tau$ must be odd with respect to CP.

While we have discussed only the important case of $K\pi$ decay, similar
considerations should hold for quasi-two-body final states with the same
strangeness
content.  It is also possible to determine $x \sin 2\phi$ from the CP-violating
asymmetry{\cite{7}} between $D^0$ and $\overline{D^0}$ decays to CP eigenstates
such
as $\pi^+ \pi^-$ or $K^+K^-$.  This is completely analogous to the proposed
experiments in $B^0$ physics{\cite{8}} except that in the $B^0$ case the mixing
is
large and has already been determined independently of any CP violation.

While completing this paper I received a preprint by Blaylock, Seiden, and
Nir{\cite{9}} emphasizing the importance of the linear term.  I am indebted to
Gustavo Branco and Ritchie Patterson for discussions and to the paper of T.
Liu{\cite{6}} that led me to these results.  This research was supported in
part by
the US DOE Contract No. DE-FG02-91ER40682.

\end{document}